\documentclass[useAMS,usenatbib]{mnras}
\usepackage{amsfonts}
\usepackage{amsmath}
\usepackage{amssymb}

\usepackage[T1]{fontenc}

\usepackage[dvips]{graphicx}
\usepackage{myaasmacros}
\usepackage{color}
\usepackage{multirow}

\def\bea{\begin{eqnarray}}
\def\ena{\end{eqnarray}}

\newcommand{\mr}[1]{\mathrm{#1}}

\title[FRB counterparts in the scenario of supergiant pulses]
{Fast Radio Bursts counterparts in
the scenario of supergiant pulses}

\author[S.~B.~Popov \&  M.~S.~Pshirkov]{S.~B.~Popov$^{1}$\thanks{E-mail: polar@sai.msu.ru}, M.~S.~Pshirkov$^{1,2,3}$\thanks{E-mail: pshirkov@sai.msu.ru}\\
$^{1}$Sternberg Astronomical Institute, Lomonosov Moscow State University, Universitetsky prospekt 13, 119992, Moscow, Russia\\
$^{2}$Pushchino Radio Astronomy Observatory, 142290 Pushchino, Russia\\
$^{3}$Institute for Nuclear Research of the Russian Academy of Sciences, 117312, Moscow, Russia\\
}

\begin{document}

\date{}

\pubyear{2016}

\maketitle

\label{firstpage}

\begin{abstract}
 We discuss identification of
possible counterparts and persistent sources related to Fast
Radio Bursts (FRBs) in the framework of the model of supergiant pulses from young
neutron stars with large spin-down luminosities.  
In particular, we demonstrate that at least some of sources of FRBs can be
observed as ultraluminous X-ray sources (ULXs).
At the moment no ULXs are known to be coincident with localization areas of
FRBs. We searched for a  correlation of FRB positions with  galaxies  in the
2MASS Redshift survey catalogue. Our analysis  produced 
statistically insignificant overabundance 
($p$-value~$\approx 4\%$)
of galaxies in error
boxes of FRBs. 
In the very near future with even modestly increased statistics of FRBs and  
with the help of dedicated X-ray observations 
and all-sky X-ray surveys it will be possible to decisively prove
or falsify the supergiant pulses model. 
\end{abstract}


\section{Introduction}
\label{sec:intro}

Fast radio bursts (FRBs) comprise  a new emerging class of
radio transients (see a review in \citealt{2016arXiv160401799K}).  
At the moment, 17 sources are known (see the catalogue in
\citealt{2016arXiv160103547P}).
They are characterised by short durations ($\sim$~few~msec),
and large values of dispersion measure (DM) which are much larger than the
expected Galactic contribution \citep{NE2001}.  If these large values were
obtained during propagation through extragalactic medium, this would firmly
put FRBs at cosmological distances, $d > 1$~Gpc, and would correspond to
gigantic energy outputs (just in radio waves!) of these events:

\begin{equation}
 L_{\mr{r}}\sim 10^{42}\left(\frac{S_\mr{peak}}{\mathrm{Jy}}\right)
\left(\frac{\Delta\nu}{1.4\, \mathrm{GHz}}\right)
\left(\frac{d}{1~\mr{Gpc}}\right)^{2}~\mr{erg~s^{-1}}.
 \label{eq:frb_lum}
 \end{equation}
\begin{equation}
 E_{\mr{r}}\sim 10^{39}\left(\frac{\tau}{1~\mr{msec}}\right)
\left(\frac{d}{1~\mr{Gpc}}\right)^{2}\left(\frac{L_{\mr{iso}}}{10^{42}~\mr{erg~s^{-1}}}\right)~\mr{erg}.
 \label{eq:frb_energy}
 \end{equation}
Here $d$ is the distance to the source, $L_{\mr{r}}$ and $E_{\mr{r}}$ are the 
radio luminosity and energy output under assumption of isotropy correspondingly,
$S_\mr{peak}$ --
peak flux, $\tau$ is the duration of the burst, and  $\Delta \nu$ is the  range of
frequencies which for estimates we set equal to the typical frequency of
observation of FRBs --- 1.4 GHz. 
 
 Short duration of these events implies that size of the active region 
is very small, $ \lesssim 10^{8}$~cm, 
which makes neutron stars-related phenomena
the most plausible candidates for explanation. 
Still, the FRBs could be a heterogeneous phenomena, 
consisting of several sub-populations for which different mechanisms of
burst emission might be applied.
 
Many models have been proposed to explain the nature of FRBs (see, for
example, references in
\citealt{2016arXiv160401799K}). Naturally, the majority of these scenarios are related to
neutron stars (NSs).
Among them several broad categories can be distinguished:
 \begin{itemize}
\item  FRBs are due to collapse of a supramassive NS to a black hole
(\citealt{2014A&A...562A.137F}).
  \item FRBs are generated  during NS-NS mergers
(\citealt{2010Ap&SS.330...13P,2013PASJ...65L..12T}). 
  \item FRBs are produced in (or after) magnetar hyperflares
(\citealt{2010vaoa.conf..129P, 2014MNRAS.442L...9L})
  \item FRBs are phenomena akin to the Crab giant pulses (GPs). Very young  
fast-rotating pulsars (PSRs) with ages less than $\sim$100 
years can potentially demonstrate analogues of 
  GPs which are $10^{4}-10^{5}$ more luminous than in the Crab pulsar
(\citealt{2016MNRAS.457..232C, 2016MNRAS.458L..19C}). 
Such hypothetical events are dubbed ``supergiant pulses''.
These flares could be observed as FRBs on Earth. In this scenario, most of DM 
is accumulated in the very vicinity of  pulsar in its supernova remnant, 
and that allow to put FRBs at somewhat smaller distances: 
$d \lesssim 100-200$~Mpc.  All other models assume cosmological distances
$\gtrsim1$~Gpc.
 \end{itemize}

The first among mentioned scenarios  can hardly provide a reasonable estimate
for the rate of events inferred from the observations. The energy output is highly uncertain.
In addition, as FRBs are not shown to be coincident
with supernova (SN).  Altogether, this means that
the model meets some severe restrictions.

The model with coalescing NSs could easily meet necessary energetic 
requirements, but also 
have serious difficulties explaining  rate of 
FRBs and repetitive bursts.

At the end of 2015 the magnetar model was considered as nearly the best, 
but still any confirmations based on observations of Galactic magnetars 
are lacking (i.e., up to date 
there are no detections of radio bursts coincident with high energy flares;
see \citealt{2016arXiv160202188T} 
and discussion in \citealt{2016arXiv160401799K}).

So, below we mainly focus on the model of supergiant pulses.\footnote{When
this paper was ready for submission, Lyutikov and Lorimer submitted an
e-print (arXiv: 1605.01468) in which they addressed the question of
contemporaneous counterparts of FRBs in the
framework of a magnetar flare.} 
In this note we will briefly 
analyse potentially testable predictions for multiwavelength observations
of  the sources of FRBs in this scenario.

\section{FRBs by energetic radio pulsars}
\label{sec:lbp2016}

\cite{2016arXiv160302891L} developed further the model in which FRBs are due
to supergiants pulses of PSRs (\citealt{2016MNRAS.458L..19C,
2016MNRAS.457..232C}). In this section we briefly describe the main features
of this scenario.

In this model, a FRB is emitted by a very energetic PSR. Expected
spin-down luminosity, $\dot E$, are $\sim10^{43}$~erg~s$^{-1}$.  The emission
mechanism is supposed to be similar to the mechanism of GPs, but
the maximal luminosity is scaled linearly with $\dot E$ (note, that FRBs can
be longer than GPs; as FRBs are widened due to scattering, and it is
difficult to derive their intrinsic duration. So, scaling of the total
energy release is a more complicated subject). 
Then, it is possible to
obtain radio pulses $\sim 10^5$ stronger than GPs of the Crab. Such
events might explain properties (peak fluxes)
of known FRBs, if observed from 100-200 Mpc.
 

These distances guarantee roughly isotropic distribution of sources in the
sky. 
At the moment the observational data are in an agreement  with such
isotropy. Still, we note, that unless the statistics is significantly higher,
all analyses of isotropy are strongly limited by a small number of known
sources. If sources are indeed inside $\sim200$~Mpc radius sphere, then it can be possible
in the near future to probe deviation from the isotropy due to still slightly inhomogeneous distribution of galaxies in this volume (see, for example,
\citealt{2001MNRAS.328.1039C}).
 
Mostly, the DM is due
to a still dense shell (supernova remnant) around the NS. Then, expected
ages of such PSRs are about few tens of years. 
For estimates below we use as a typical value
the age 30 years.


Note, that as in the model with supergiant pulses 
and in the model with magnetar flares FRBs are
related to young NS, sources might be located in regions of intense
starformation. Then, significant DM can be partially due to the interstellar medium in
the local surroundings. In any case, absence of FRBs with low DM requires some
``guaranteed'' DM, either due to a SNR, or due to intergalactic medium.

Calculations (see \citealt{2016arXiv160302891L}) show that 
it is possible to explain the estimated FRBs rate $\sim10^4$ per day 
by the population 
of young ($\lesssim$ few tens of years) energetic PSRs within 100-200 Mpc
from the Sun assuming that the repetition rate is $\lesssim 1$ per day,
in correspondence with observations.
This estimate is based on  the core-collapse SN rate
$\sim 3 \times 10^{-4}$~yr$^{-1}$~Mpc$^{-3}$
(\citealt{2012ApJ...757...70D}).

The supergiant pulses 
scenario predicts that FRBs should
repeat quite frequently.  Given that all FRBs were observed with high
signal-to-noise ratio \citep{2016arXiv160103547P},
 and that GP rate falls very quickly with
increasing $L_\mr{r}$, it is natural to expect fainter but much more frequent
repetitive FRBs. The  FRB 121102 can be an example of such
behaviour \citep{2016Natur.531..202S}. 

We can expect to have $\gtrsim 10^5$ large  galaxies within this volume. 
This corresponds to about one source per 10 galaxies. Near-by
population of galaxies is relatively well studied, and if sources of FRBs
remain bright at some energy range even between the bursts, then we can hope
to identify them in catalogues, archival data, or dedicated observations.

Young PSRs with large $\dot E$ are known to be bright X-ray sources.
According to \cite{2002A&A...387..993P}
X-ray luminosity of such a source can be estimated as:
$L_\mathrm{X}=10^{-15.3} (\dot E)^{1.34}$.  In addition, some fraction of 
total energy losses might be re-emitted by a pulsar wind nebula (PWN). 

A shell around the PSR relatively quickly, --- within few years, ---
 becomes transparent for X-rays
(\citealt{2016arXiv160308875M}). So, at the ages required in the scenario by
\cite{2016arXiv160302891L} a bright (possibly ultraluminous) X-ray source might be
observed.

In this respect, 
what would be the observational consequences of the supergiant PSR burst 
scenario? How 
presence of several thousand energetic
($\dot E\sim10^{43}~\mr{erg~s^{-1}}$) in 200~Mpc radius sphere (given
100\% fraction of young pulsar-related FRBs) can be probed?
We suggest that it might result in appearance of many ultraluminous X-ray
sources (ULXs). This conclusion seems to be unavoidable in this framework,
and we discuss it in the following section.

\section{ULXs as conterparts of young energetic pulsars}
\label{sec:ulx}

A young PSR might have an X-ray luminosity:
 
$$
L_\mathrm{X}\approx 2 \times 10^{42} \left(\dot E/10^{43}~\mr{erg~s}^{-1} 
\right)^{1.34}~\mr{erg~s}~^{-1},$$ 
(see \citealt{2002A&A...387..993P}). 
However, this relation is not probed for
very high values of $\dot E$, and so its usage is just an extrapolation.
Most probably, $L_\mr{X}$ does not grow that fast with $\dot E$ for large
values (still, for sure we expect to have a very bright X-ray sources for
large rotational energy losses). 
On the other hand, significant additional
X-ray emission can appear due to a PWN (\citealt{2013arXiv1305.2552K}).
\cite{2004ApJ...615..222P} suggested that some of ultraluminous X-ray
sources (ULXs)
can be young energetic PSRs.  Later, \cite{2013MNRAS.431.2690M} developed
this idea.
Such sources might dominate over high-mass X-ray binaries (HMXB) at high
($L_\mr{X}>10^{40}~\mr{erg~s^{-1}}$) luminosities 
\citep{2004ApJ...615..222P,
2013MNRAS.431.2690M}.  

\cite{2011ApJ...741...49S} studied a large (nearly complete, if we speak
about large bright objects) sample of galaxies within 14.5 Mpc.  
They identified
$\sim100$ ULXs in 124 galaxies with total starformation rate (SFR) $\sim 50~
M_\odot$~yr$^{-1}$, but  
found no sources with luminosities
$L_\mr{X}>10^{41}~\mr{erg~s^{-1}}$ in their sample.  
Nevertheless, it does not
contradict the possibility that there could be $N_{200}\sim10^{4}$ such
sources in 200~Mpc radius sphere around the Sun.  
For the local starformation rate density
value $(0.022\pm0.001)~M_{\odot}~\rm{yr^{-1}~Mpc^{-3}}$
\citep{2007MNRAS.375..931M} it is expected that only
$0.67\left({N_{200}}/{10^{4}}\right)$ sources with such luminosity might be 
found in the
aforementioned survey. 
Then non-detection of bright ULXs by \cite{2011ApJ...741...49S} does not
invalidate the scenario. 

The most natural consequence  of the model --- 
which holds even if there was no
ULX-FRB connection --- is an expected considerable positional correlation of
FRBs with nearby galaxies. 
Unfortunately, limited angular
resolution of FRB localization 
($\sim 14$-$15$ arcmin in the case of Parkes telescope observations, where most of
the bursts have been identified, see \citealt{2016arXiv160103547P}) and
scarce statistics significantly complicate
deriving any definite conclusions.  

Nevertheless, we
correlated FRB positions with bright galaxies from the 2MASS Redshift Survey
(2MRS, \citealt{2012ApJS..199...26H}).  There are 37 209 galaxies with radial
velocities smaller than 14 000 $\mr{km~s^{-1}}$ which corresponds to
$d<200$~Mpc in this survey.  It is expected to have 2.27 coincidences in
circles with radius $15$~arcmin\footnote{We use a more conservative estimate
for the positional accuracy which is double of the FWHM of a Parkes beam.} 
for 17 FRB by chance alone, 
and we found 5, which gives a
Poissonian probability $p\sim8\%$.  More than that, 3 out of 17 FRBs 
(FRB 010621, FRB 121102, and FRB 150418) are
very close to the Galactic plane and fall inside the ``blind spot'' of the
2MRS catalogue -- it was constructed using sources with $|b|>5^{{\circ}}$
($|b|>8^{{\circ}}$ closer to the bulge).  This fact lowers probability to
$4\%$ level.  Nevertheless, given unknown exposure map of the current FRB
catalogue it is difficult to make any far-reaching claims.  However, if this
$\sim2\sigma$ fluctuation was real, then a meagre trebling of the number of
detected FRBs would be enough for FRB-2MRS correlation to reach a high level
of statistical significance.

Young energetic pulsar might demonstrate Crab-like properties, 
i.e.  they might be surrounded by a very bright PWN, which can
be used as a multiwavelength beacon.  Huge spin-down luminosity cannot just
disappear without a trace (for example, carried away by a relativistic
particle flow). 
Instead it must be (partially) dissipated and
radiated in different energy ranges.  

In addition, 
these objects should be bright persistent radio sources, 
so additional cross-correlation between X-ray and radio catalogues
could lead to interesting results.  

\section{Future searches for persistent counterparts}
\label{sec:disc}

In this section we discuss several issues related to the FRB-ULX connection.

\subsection{Possible \textit{XMM-Newton} observations}
\label{sec:disc_xmm}

Identification of host galaxies of FRBs is complicated as their positions
are not well-known. Most of the bursts have been discovered at the Parkes 
telescope. Then, the uncertainty in position is about the size of the beam
of Parkes. Full beam width on half maximum amplitude is
$\approx 14$-15 arcmin (\citealt{2016arXiv160103547P}). 
Note, that the field of view of the EPIC instrument onboard \textit{XMM-Newton} is 
about 30$\arcmin$ in diameter(\citealt{Turner:2000jy}). If FRBs are mostly (or at least partly) due to
supergiant pulses of energetic PSRs, then we can expect to find an X-ray
source with flux $f\approx8\times10^{-13} (L_\mr{X}/10^{42}~\mr{erg~s}^{-1})
d_{100}^{-2}$~erg~cm$^{-2}$~s$^{-1}$.  

In the framework of the scenario developed by \cite{2016arXiv160302891L} 
FRB luminosity is ~$L_\mathrm{r}=\eta \dot E$, where $\eta\approx 0.01$.
FRBs with different observed peak
fluxes are nearly uniformly distributed in distances, and DM is a poor
indicator of distance to the source. 
According to \cite{2002A&A...387..993P}
$L_\mr{X}\sim\dot E^{1.34}$.   
Then it is possible to find a relatively
bright (in terms of the flux) ULX coincident with a bright FRB in a relatively
distant galaxy. Moreover, if peak flux of FRBs is not correlated with the
distance (or correlates very weakly), then brighter (in terms of flux) ULXs
can be found in more distant galaxies.  
This can be illustrated as follows. 


 For a typical FRB with peak flux $S_\mr{peak}=$~1~Jy we obtain radio
luminosity:

$$L_\mr{r}=1.7\times 10^{40} (S_\mr{peak}/1\, \mr{Jy})
(d/100\, \mathrm{Mpc})^2 \, \mathrm{erg} \, \mathrm{s}^{-1}.$$ 
Then, rotational energy losses are:

$$\dot E=1.7\times 10^{42} (S_\mr{peak}/1\, \mathrm{Jy})
(d/100\, \mathrm{Mpc})^2(\eta/0.01)^{-1}\,\mathrm{erg}\,\mathrm{s}^{-1}.$$ 
Using the relation from Possenti et al. we obtain the X-ray luminosity:

$$L_\mr{X}=1.8\times 10^{41} (S_\mr{peak}/1\, \mr{Jy})^{1.34} \times
$$
$$
\times (d/100\, \mathrm{Mpc})^{2.68} (\eta/0.01)^{-1.34}\, \mathrm{erg}\,\mathrm{s}^{-1}.$$
And so, the X-ray flux is:

 $$f_\mr{X}=1.5\times 10^{-13} (S_\mr{peak}/1 
\, \mr{Jy})^{1.34} \times
$$
$$ \times (d/100\, \mathrm{Mpc})^{0.68} (\eta/0.01)^{-1.34}
\,~\mathrm{erg}\,~\mathrm{cm}^{-2}~\, \mathrm{s}^{-1}.$$ 
For large distances we obtain higher $f_\mr{X}$ for a given $S_\mr{peak}$,
for smaller --- weaker. 
If a source with peak flux 1 Jy is at 10 Mpc, then $f_\mr{X}=3.2\times
10^{-14}$~erg~cm$^{-2}$~s$^{-1}$. Correspondently, for 200 Mpc we have
$f_\mr{X}=2.5\times 10^{-13}$~erg~cm$^{-2}$~s$^{-1}$.

Then, a limit is determined by closer sources.\footnote{Note, that the
applied relation $L_\mr{X} vs. \dot E$ can not be valid for large luminosities.}
Non-detection of a stable X-ray source down to
$f\sim10^{-14}$~--~$10^{-15}$~~erg~cm$^{-2}$~s$^{-1}$ can be a useful constraint.

\subsection{Search for FRBs in ``supernova factories''}
\label{sec:disc_snf}

Inside $\lesssim 50$~Mpc there are few galaxies with extreme values of
starformation rate and supernova rate. In particular, we note Arp 299
(\citealt{2004ApJ...611..186N}), 
and NGC 3256 (\citealt{2004MNRAS.354L...1L}).

The rate of SNae in these galaxies is $\sim 1$ per year. Then, even if just
1/10 of young pulsars are energetic enough to produce a detectable FRB from
$d\sim40$~Mpc, then we can expect several of such sources in each galaxy. With
the rate of repetition $\sim$ once per day, it is worth trying to monitor
these galaxies in radio. Identification of an ULX, however, would be very
problematic in such cases due to crowding of X-ray sources.

\subsection{Sources in local starforming galaxies}
\label{sec:disc_sfg}

Inside $\lesssim 4$~Mpc most of starformation is related to just 
four galaxies:
M82, M83, NGC 253, NGC 4945 (\citealt{1998ASPC..148..127H}). 
Typically, SN rate in each of 
these galaxies is higher than in the Milky way by a
factor of a few (up to 10, see data and references in
\citealt{2005astro.ph..2391P}).  

We can expect few PSRs with ages $\lesssim 30$~yrs in each of these galaxies.
Some of them can be energetic enough to produce detectable FRBs.
As galaxies are near-by, identification of ULXs would not be a very
difficult task.  

Note, that no ULXs with $L_\mr{X}>10^{41}$~erg~s$^{-1}$ have been found in
these galaxies. Then we can suspect that potential sources have smaller
$\dot E$, and so produce radio bursts with luminosities smaller that those
of classical FRBs. Still, radio monitoring is worthwhile due to proximity of
these galaxies. 

\subsection{Bursts from M31}
\label{sec:disc_m31}

M31 is the closest large galaxy.
Recently 
\cite{2013MNRAS.428.2857R} reported discovery of several millisecond 
radio bursts from it. No periodicity have been found (so, the
interpretation based on radio pulsars or RRATs is not viable), however, some
sources can be repititive. \cite{2013arXiv1307.4924P} suggested that this
flares can be weak relatives of FRBs, originating from the same type of
sources which demonstrate activity in different ranges of released energy
(in the particular model these two types of activity are hyperflares of
magnetars and their usual weak bursts). 

A similar interpretation can be made
in the case of supergiant pulses. I.e., we observed analogues of FRBs from
more numerous PSRs with smaller $\dot E$, which cannot produce strong bursts.
Then, search for weak FRBs from local galaxies 
can be fruitful (or can put important constraints on the model
of supergiant pulses).


\subsection{Future observations}

It is expected that statistics of FRBs can be greatly increased in 1-2
years \citep{2016arXiv160207292C, 2016arXiv160205165K}. 
This might be due to several new instruments. 
UTMOST is already working \citep{2016MNRAS.458..718C}, and it is expected
that it is going to contribute to the increase of the FRB statistics with
the detection rate $\sim 1$ per 1-2 weeks. 
Another telescope --- CHIME \citep{2014SPIE.9145E..22B} --- is expected to
start gathering data very soon and reach the rate up to $\sim 1$ per day, if
at lower frequencies FRBs are well-visible.  

In the fall of 2016 the Five hundred meter Aperture Spherical Telescope
(FAST, \citealt{2011IJMPD..20..989N}) might be completed. 
It is expected that this instrument
will detect one new FRB in a week of operation \citep{2016arXiv160206099L}.

In not-so-close future SKA  will become extremely effective, detecting
nearly a FRB each hour in the final configuration
\citep{2015aska.confE..55M, 2016arXiv160205165K}! 

By itself, new radio data can be used to probe many proposed models of FRBs.
For example, in the scenario with supergiant pulses we expect that with
$\sim100$ sources we can easily reach a statistically significant level of correlation with local galaxies. 
New radio observations can be complemented by a new sensitive all-sky X-ray
survey by eROSITA onboard \textit{Spektrum-Roentgen-Gamma} \citep{2011SPIE.8145E..0DP}. 
This would make testing this model even easier. And if the model is correct,
than we can expect many associations of FRBs with ULXs due to new
observations.


\section{Summary}\label{sec:summary}
The supergiant pulses model of FRBs can be tested on the base of  
a direct identification of sources, because in this framework
they are young energetic pulsars
residing quite close to us, $d<200$~Mpc.  
Large spin-down
luminosity of these pulsarss, $\dot E\sim10^{43}~\mr{erg~s^{-1}}$, will
lead to emergence of bright counterparts at various frequencies.  The
pulsars (and also, possibly, their PWNs, see
\citealt{2013arXiv1305.2552K}) might be luminous X-ray sources and eventually
can manifest themselves as ULXs with luminosities that can even overcome
the brightest HMXBs, $L_\mr{X}>10^{41}$~erg~s$^{-1}$.  
Unfortunately, at the moment no FRB 
are known in the regions observed by the \textit{XMM-Newton} X-ray
observatory.  There are two natural avenues to pursue: first, dedicated
observations at several directions, coinciding with FRB localizations, can
be performed; second, one can search for unusual ULXs in archival data.
  
As already stated above, 
FRBs can also be rather bright persistent sources in radio waveband,
and this is crucial for discrimination between young pulsars and
HMXBs\footnote{Given that we are not dealing with the extreme case of ULX
with $L_\mr{X}>10^{41}$, as HMXBs with such luminosity might be extremely
rare, or even absent.}: one will not
expect any sizeable radio-emission from HMXBs, beside rare cases of
microquasars, which can be mostly filtered out due to their variability.  
It can
also be the case when we are trying to discriminate against background AGNs
that can mimic our sought sources.  Also, after accurate pin-pointing of
candidate position with X-ray and radio observations, it is possible to
search for counterparts at other frequiencies -- in optics, IR, or UV.

Finally, the conclusion that the FRBs should be local phenomena
($d<200$~Mpc) can be tested even if these sources are relatively
underluminous, because then a significant positional correlation with nearby
galaxies is expected.  We correlated FRB positions with bright galaxies from
the 2MRS catalogue. We found 5 pairs FRB-galaxy 
with distance less than 15$\arcmin$,
and 1.87 coincidences were expected by chance, giving a Poissonian
probability $p\sim4\%$.  With even modest increase in total FRB number the
fraction of local population will be estimated (or, seriously constrained)
in the very near future.

\section*{Acknowledgements}
The work of  M.S.P. is supported by RSF grant No. 14-12-00146.
S.B.P. acknowledges support from RFBR (project 14-02-00657).  
This research has made use of NASA's Astrophysics Data System.
The authors thank Prof. K.A. Postnov and Dr. I. Zolotukhin for discussions.
Also S.B.P. thanks for discussions Prof. M. Lyutikov.
S.B.P. is the ``Dynasty'' Foundation fellow.




\bibliographystyle{mnras}
\bibliography{frb,ulx}

\begin{thebibliography}{}
\makeatletter
\relax
\def\mn@urlcharsother{\let\do\@makeother \do\$\do\&\do\#\do\^\do\_\do\%\do\~}
\def\mn@doi{\begingroup\mn@urlcharsother \@ifnextchar [ {\mn@doi@}
  {\mn@doi@[]}}
\def\mn@doi@[#1]#2{\def\@tempa{#1}\ifx\@tempa\@empty \href
  {http://dx.doi.org/#2} {doi:#2}\else \href {http://dx.doi.org/#2} {#1}\fi
  \endgroup}
\def\mn@eprint#1#2{\mn@eprint@#1:#2::\@nil}
\def\mn@eprint@arXiv#1{\href {http://arxiv.org/abs/#1} {{\tt arXiv:#1}}}
\def\mn@eprint@dblp#1{\href {http://dblp.uni-trier.de/rec/bibtex/#1.xml}
  {dblp:#1}}
\def\mn@eprint@#1:#2:#3:#4\@nil{\def\@tempa {#1}\def\@tempb {#2}\def\@tempc
  {#3}\ifx \@tempc \@empty \let \@tempc \@tempb \let \@tempb \@tempa \fi \ifx
  \@tempb \@empty \def\@tempb {arXiv}\fi \@ifundefined
  {mn@eprint@\@tempb}{\@tempb:\@tempc}{\expandafter \expandafter \csname
  mn@eprint@\@tempb\endcsname \expandafter{\@tempc}}}

\bibitem[\protect\citeauthoryear{{Bandura} et~al.,}{{Bandura}
  et~al.}{2014}]{2014SPIE.9145E..22B}
{Bandura} K.,  et~al., 2014, in Ground-based and Airborne Telescopes V. p.
  914522 (\mn@eprint {arXiv} {1406.2288}), \mn@doi{10.1117/12.2054950}

\bibitem[\protect\citeauthoryear{{Caleb} et~al.,}{{Caleb}
  et~al.}{2016}]{2016MNRAS.458..718C}
{Caleb} M.,  et~al., 2016, \mn@doi [\mnras] {10.1093/mnras/stw109}, \href
  {http://adsabs.harvard.edu/abs/2016MNRAS.458..718C} {458, 718}

\bibitem[\protect\citeauthoryear{{Colless} et~al.,}{{Colless}
  et~al.}{2001}]{2001MNRAS.328.1039C}
{Colless} M.,  et~al., 2001, \mn@doi [\mnras]
  {10.1046/j.1365-8711.2001.04902.x}, \href
  {http://adsabs.harvard.edu/abs/2001MNRAS.328.1039C} {328, 1039}

\bibitem[\protect\citeauthoryear{{Connor}, {Lin}, {Masui}, {Oppermann}, {Pen},
  {Peterson}, {Roman}  \& {Sievers}}{{Connor}
  et~al.}{2016a}]{2016arXiv160207292C}
{Connor} L.,  {Lin} H.-H.,  {Masui} K.,  {Oppermann} N.,  {Pen} U.-L.,
  {Peterson} J.~B.,  {Roman} A.,   {Sievers} J.,  2016a, preprint, \href
  {http://adsabs.harvard.edu/abs/2016arXiv160207292C} {} (\mn@eprint {arXiv}
  {1602.07292})

\bibitem[\protect\citeauthoryear{{Connor}, {Sievers}  \& {Pen}}{{Connor}
  et~al.}{2016b}]{2016MNRAS.458L..19C}
{Connor} L.,  {Sievers} J.,   {Pen} U.-L.,  2016b, \mn@doi [\mnras]
  {10.1093/mnrasl/slv124}, \href
  {http://adsabs.harvard.edu/abs/2016MNRAS.458L..19C} {458, L19}

\bibitem[\protect\citeauthoryear{{Cordes} \& {Lazio}}{{Cordes} \&
  {Lazio}}{2002}]{NE2001}
{Cordes} J.~M.,  {Lazio} T.~J.~W.,  2002, ArXiv: astro-ph/0207156, \href
  {http://adsabs.harvard.edu/abs/2002astro.ph..7156C} {}

\bibitem[\protect\citeauthoryear{{Cordes} \& {Wasserman}}{{Cordes} \&
  {Wasserman}}{2016}]{2016MNRAS.457..232C}
{Cordes} J.~M.,  {Wasserman} I.,  2016, \mn@doi [\mnras]
  {10.1093/mnras/stv2948}, \href
  {http://adsabs.harvard.edu/abs/2016MNRAS.457..232C} {457, 232}

\bibitem[\protect\citeauthoryear{{Dahlen}, {Strolger}, {Riess}, {Mattila},
  {Kankare}  \& {Mobasher}}{{Dahlen} et~al.}{2012}]{2012ApJ...757...70D}
{Dahlen} T.,  {Strolger} L.-G.,  {Riess} A.~G.,  {Mattila} S.,  {Kankare} E.,
  {Mobasher} B.,  2012, \mn@doi [\apj] {10.1088/0004-637X/757/1/70}, \href
  {http://adsabs.harvard.edu/abs/2012ApJ...757...70D} {757, 70}

\bibitem[\protect\citeauthoryear{{Falcke} \& {Rezzolla}}{{Falcke} \&
  {Rezzolla}}{2014}]{2014A&A...562A.137F}
{Falcke} H.,  {Rezzolla} L.,  2014, \mn@doi [\aap]
  {10.1051/0004-6361/201321996}, \href
  {http://adsabs.harvard.edu/abs/2014A%26A...562A.137F} {562, A137}

\bibitem[\protect\citeauthoryear{{Heckman}}{{Heckman}}{1998}]{1998ASPC..148..127H}
{Heckman} T.~M.,  1998, in {Woodward} C.~E.,  {Shull} J.~M.,   {Thronson} Jr.
  H.~A.,  eds,  Astronomical Society of the Pacific Conference Series Vol. 148,
  Origins. p.~127 (\mn@eprint {} {astro-ph/9708263})

\bibitem[\protect\citeauthoryear{{Huchra} et~al.,}{{Huchra}
  et~al.}{2012}]{2012ApJS..199...26H}
{Huchra} J.~P.,  et~al., 2012, \mn@doi [\apjs] {10.1088/0067-0049/199/2/26},
  \href {http://adsabs.harvard.edu/abs/2012ApJS..199...26H} {199, 26}

\bibitem[\protect\citeauthoryear{{Kargaltsev}, {Rangelov}  \&
  {Pavlov}}{{Kargaltsev} et~al.}{2013}]{2013arXiv1305.2552K}
{Kargaltsev} O.,  {Rangelov} B.,   {Pavlov} G.~G.,  2013, preprint, \href
  {http://adsabs.harvard.edu/abs/2013arXiv1305.2552K} {} (\mn@eprint {arXiv}
  {1305.2552})

\bibitem[\protect\citeauthoryear{{Katz}}{{Katz}}{2016}]{2016arXiv160401799K}
{Katz} J.~I.,  2016, preprint, \href
  {http://adsabs.harvard.edu/abs/2016arXiv160401799K} {} (\mn@eprint {arXiv}
  {1604.01799})

\bibitem[\protect\citeauthoryear{{Keane} \& {SUPERB Collaboration}}{{Keane} \&
  {SUPERB Collaboration}}{2016}]{2016arXiv160205165K}
{Keane} E.~F.,  {SUPERB Collaboration} 2016, preprint, \href
  {http://adsabs.harvard.edu/abs/2016arXiv160205165K} {} (\mn@eprint {arXiv}
  {1602.05165})

\bibitem[\protect\citeauthoryear{{Li}, {Huang}, {Zhang}, {Li}  \& {Li}}{{Li}
  et~al.}{2016}]{2016arXiv160206099L}
{Li} L.,  {Huang} Y.,  {Zhang} Z.,  {Li} D.,   {Li} B.,  2016, preprint, \href
  {http://adsabs.harvard.edu/abs/2016arXiv160206099L} {} (\mn@eprint {arXiv}
  {1602.06099})

\bibitem[\protect\citeauthoryear{{L{\'{\i}}pari} et~al.,}{{L{\'{\i}}pari}
  et~al.}{2004}]{2004MNRAS.354L...1L}
{L{\'{\i}}pari} S.~L.,  et~al., 2004, \mn@doi [\mnras]
  {10.1111/j.1365-2966.2004.08213.x}, \href
  {http://adsabs.harvard.edu/abs/2004MNRAS.354L...1L} {354, L1}

\bibitem[\protect\citeauthoryear{{Lyubarsky}}{{Lyubarsky}}{2014}]{2014MNRAS.442L...9L}
{Lyubarsky} Y.,  2014, \mn@doi [\mnras] {10.1093/mnrasl/slu046}, \href
  {http://adsabs.harvard.edu/abs/2014MNRAS.442L...9L} {442, L9}

\bibitem[\protect\citeauthoryear{{Lyutikov}, {Burzawa}  \& {Popov}}{{Lyutikov}
  et~al.}{2016}]{2016arXiv160302891L}
{Lyutikov} M.,  {Burzawa} L.,   {Popov} S.~B.,  2016, preprint, \href
  {http://adsabs.harvard.edu/abs/2016arXiv160302891L} {} (\mn@eprint {arXiv}
  {1603.02891})

\bibitem[\protect\citeauthoryear{{Macquart} et~al.,}{{Macquart}
  et~al.}{2015}]{2015aska.confE..55M}
{Macquart} J.~P.,  et~al., 2015, Advancing Astrophysics with the Square
  Kilometre Array (AASKA14), \href
  {http://adsabs.harvard.edu/abs/2015aska.confE..55M} {p.~55}

\bibitem[\protect\citeauthoryear{{Mauch} \& {Sadler}}{{Mauch} \&
  {Sadler}}{2007}]{2007MNRAS.375..931M}
{Mauch} T.,  {Sadler} E.~M.,  2007, \mn@doi [\mnras]
  {10.1111/j.1365-2966.2006.11353.x}, \href
  {http://adsabs.harvard.edu/abs/2007MNRAS.375..931M} {375, 931}

\bibitem[\protect\citeauthoryear{{Medvedev} \& {Poutanen}}{{Medvedev} \&
  {Poutanen}}{2013}]{2013MNRAS.431.2690M}
{Medvedev} A.~S.,  {Poutanen} J.,  2013, \mn@doi [\mnras]
  {10.1093/mnras/stt369}, \href
  {http://adsabs.harvard.edu/abs/2013MNRAS.431.2690M} {431, 2690}

\bibitem[\protect\citeauthoryear{{Murase}, {Kashiyama}  \& {Meszaros}}{{Murase}
  et~al.}{2016}]{2016arXiv160308875M}
{Murase} K.,  {Kashiyama} K.,   {Meszaros} P.,  2016, preprint, \href
  {http://adsabs.harvard.edu/abs/2016arXiv160308875M} {} (\mn@eprint {arXiv}
  {1603.08875})

\bibitem[\protect\citeauthoryear{{Nan} et~al.,}{{Nan}
  et~al.}{2011}]{2011IJMPD..20..989N}
{Nan} R.,  et~al., 2011, \mn@doi [International Journal of Modern Physics D]
  {10.1142/S0218271811019335}, \href
  {http://adsabs.harvard.edu/abs/2011IJMPD..20..989N} {20, 989}

\bibitem[\protect\citeauthoryear{{Neff}, {Ulvestad}  \& {Teng}}{{Neff}
  et~al.}{2004}]{2004ApJ...611..186N}
{Neff} S.~G.,  {Ulvestad} J.~S.,   {Teng} S.~H.,  2004, \mn@doi [\apj]
  {10.1086/383608}, \href {http://adsabs.harvard.edu/abs/2004ApJ...611..186N}
  {611, 186}

\bibitem[\protect\citeauthoryear{{Perna} \& {Stella}}{{Perna} \&
  {Stella}}{2004}]{2004ApJ...615..222P}
{Perna} R.,  {Stella} L.,  2004, \mn@doi [\apj] {10.1086/423950}, \href
  {http://adsabs.harvard.edu/abs/2004ApJ...615..222P} {615, 222}

\bibitem[\protect\citeauthoryear{{Petroff} et~al.,}{{Petroff}
  et~al.}{2016}]{2016arXiv160103547P}
{Petroff} E.,  et~al., 2016, preprint, \href
  {http://adsabs.harvard.edu/abs/2016arXiv160103547P} {} (\mn@eprint {arXiv}
  {1601.03547})

\bibitem[\protect\citeauthoryear{{Popov}}{{Popov}}{2005}]{2005astro.ph..2391P}
{Popov} S.~B.,  2005, preprint, \href
  {http://adsabs.harvard.edu/abs/2005astro.ph..2391P} {} (\mn@eprint {}
  {astro-ph/0502391})

\bibitem[\protect\citeauthoryear{{Popov} \& {Postnov}}{{Popov} \&
  {Postnov}}{2010}]{2010vaoa.conf..129P}
{Popov} S.~B.,  {Postnov} K.~A.,  2010, in {Harutyunian} H.~A.,  {Mickaelian}
  A.~M.,   {Terzian} Y.,  eds, Evolution of Cosmic Objects through their
  Physical Activity. pp 129--132 (\mn@eprint {arXiv} {0710.2006})

\bibitem[\protect\citeauthoryear{{Popov} \& {Postnov}}{{Popov} \&
  {Postnov}}{2013}]{2013arXiv1307.4924P}
{Popov} S.~B.,  {Postnov} K.~A.,  2013, preprint, \href
  {http://adsabs.harvard.edu/abs/2013arXiv1307.4924P} {} (\mn@eprint {arXiv}
  {1307.4924})

\bibitem[\protect\citeauthoryear{{Possenti}, {Cerutti}, {Colpi}  \&
  {Mereghetti}}{{Possenti} et~al.}{2002}]{2002A&A...387..993P}
{Possenti} A.,  {Cerutti} R.,  {Colpi} M.,   {Mereghetti} S.,  2002, \mn@doi
  [\aap] {10.1051/0004-6361:20020472}, \href
  {http://adsabs.harvard.edu/abs/2002A%26A...387..993P} {387, 993}

\bibitem[\protect\citeauthoryear{{Predehl} et~al.,}{{Predehl}
  et~al.}{2011}]{2011SPIE.8145E..0DP}
{Predehl} P.,  et~al., 2011, in Society of Photo-Optical Instrumentation
  Engineers (SPIE) Conference Series. p. 81450D, \mn@doi{10.1117/12.893344}

\bibitem[\protect\citeauthoryear{{Pshirkov} \& {Postnov}}{{Pshirkov} \&
  {Postnov}}{2010}]{2010Ap&SS.330...13P}
{Pshirkov} M.~S.,  {Postnov} K.~A.,  2010, \mn@doi [\apss]
  {10.1007/s10509-010-0395-x}, \href
  {http://adsabs.harvard.edu/abs/2010Ap%26SS.330...13P} {330, 13}

\bibitem[\protect\citeauthoryear{{Rubio-Herrera}, {Stappers}, {Hessels}  \&
  {Braun}}{{Rubio-Herrera} et~al.}{2013}]{2013MNRAS.428.2857R}
{Rubio-Herrera} E.,  {Stappers} B.~W.,  {Hessels} J.~W.~T.,   {Braun} R.,
  2013, \mn@doi [\mnras] {10.1093/mnras/sts205}, \href
  {http://adsabs.harvard.edu/abs/2013MNRAS.428.2857R} {428, 2857}

\bibitem[\protect\citeauthoryear{{Spitler} et~al.,}{{Spitler}
  et~al.}{2016}]{2016Natur.531..202S}
{Spitler} L.~G.,  et~al., 2016, \mn@doi [\nat] {10.1038/nature17168}, \href
  {http://adsabs.harvard.edu/abs/2016Natur.531..202S} {531, 202}

\bibitem[\protect\citeauthoryear{{Swartz}, {Soria}, {Tennant}  \&
  {Yukita}}{{Swartz} et~al.}{2011}]{2011ApJ...741...49S}
{Swartz} D.~A.,  {Soria} R.,  {Tennant} A.~F.,   {Yukita} M.,  2011, \mn@doi
  [\apj] {10.1088/0004-637X/741/1/49}, \href
  {http://adsabs.harvard.edu/abs/2011ApJ...741...49S} {741, 49}

\bibitem[\protect\citeauthoryear{{Tendulkar}, {Kaspi}  \& {Patel}}{{Tendulkar}
  et~al.}{2016}]{2016arXiv160202188T}
{Tendulkar} S.~P.,  {Kaspi} V.~M.,   {Patel} C.,  2016, preprint, \href
  {http://adsabs.harvard.edu/abs/2016arXiv160202188T} {} (\mn@eprint {arXiv}
  {1602.02188})

\bibitem[\protect\citeauthoryear{{Totani}}{{Totani}}{2013}]{2013PASJ...65L..12T}
{Totani} T.,  2013, \mn@doi [\pasj] {10.1093/pasj/65.5.L12}, \href
  {http://adsabs.harvard.edu/abs/2013PASJ...65L..12T} {65, 12}

\bibitem[\protect\citeauthoryear{Turner et~al.}{Turner
  et~al.}{2001}]{Turner:2000jy}
Turner M. J.~L.,  et~al., 2001, \mn@doi [Astron. Astrophys.]
  {10.1051/0004-6361:20000087}, 365, L27

\makeatother
\end{thebibliography}


\end{document}